\documentclass[numbers]{elsarticle}

\usepackage[top=1.25in, bottom=1.25in, left=1.5in, right=1.5in]{geometry}
\usepackage{lineno}
\usepackage{soul}
\usepackage{color}
\usepackage{mathptmx}
\usepackage{amsfonts}
\usepackage{multicol}
\usepackage{mathrsfs}
\usepackage{tensor}
\usepackage{subfig}
\usepackage{hhline}
\usepackage{upgreek}
\usepackage{cancel}
\usepackage{ulem}
\usepackage{amssymb}
\usepackage{amsthm}
\usepackage{multirow}
\usepackage{setspace}
\usepackage[flushleft]{threeparttable}
\usepackage{makecell,booktabs}
\usepackage{tikz}
\usetikzlibrary{quotes,angles}
\usepackage[pagebackref=true,
            colorlinks=true,
            bookmarks=true,
           ]{hyperref}              
\usepackage[most]{tcolorbox}           
\usepackage{graphicx}
\usepackage{epstopdf}
\usepackage{siunitx}
\usepackage{empheq}
\usepackage{amssymb}
\usepackage{array}
\usepackage{amsmath, scalerel}
\usepackage{leftidx}

\DeclareGraphicsExtensions{.eps}
\def\onedot{$\mathsurround0pt\ldotp$}
\def\cddot{
	\mathbin{\vcenter{\baselineskip.67ex
			\hbox{\onedot}\hbox{\onedot}}%
}}

\journal{Elsevier}

\usepackage{amsmath,scalerel}

\newcommand{\tightoverset}[2]{\mathop{#2}\limits^{\vbox to -.5ex{\kern-0.75ex\hbox{$#1$}\vss}}}
\modulolinenumbers[5]

\begin{document}

\allowdisplaybreaks[4]

\begin{frontmatter}
\title{On the critical role of martensite hardening behavior in the paradox of local and global ductility in dual-phase steels }

\author[1,2]{V. Rezazadeh}
\author[1]{\corref{cor}R.H.J. Peerlings}\ead{r.h.j.peerlings@tue.nl}
\author[1]{J.P.M. Hoefnagels}
\author[1]{M.G.D. Geers}

\address[1]{Department of Mechanical Engineering, Eindhoven University of Technology (TU/e), P.O.Box 513, 5600 MB Eindhoven, The Netherlands}
\address[2]{Materials Innovation Institute (M2i), P.O.Box 5008, 2600 GA Delft, The Netherlands}
\cortext[cor]{Corresponding author.}

    \begin{abstract}
	The ductility of sheet metal is typically limited by either localized necking or by damage and fracture. A recent ductility classification refers to these failure modes respectively as "global" and "local" formability. The forming limit curve (FLC) and the uniaxial tensile test assess global formability, whereas the fracture forming limit (FFL), the true thickness fracture strain, hole expansion ratios (HER), etc. are indicators of local formability. Experimental hole expansion data in the literature for different dual-phase (DP) steel grades of similar strength and composition presents a paradox: grades which are found to be ductile in a tensile test and/or FLC show a low ductility in hole expansion, whereas other grades with a low ductility in conventional tests perform surprisingly well at cut edges. In this work, an in-depth systematic statistical analysis of idealized artificial two-phase microstructures is carried out to unravel the underlying mechanisms of the observed paradoxical trends. This is done by scaling the hardness of martensite and its volume fraction to generate virtual DP steels of the same strength but different strain hardening and mechanical phase contrast (hardness difference). The proposed micromechanical model adequately reproduces the experimentally observed trends. The results show that in DP steels, a higher global (necking-driven) ductility is obtained upon delaying martensite plasticity by increasing the mechanical contrast of the two phases. Consequently, the stress-strain distributions becomes more heterogeneous, resulting in a lower fracture limit of one of the phases or interfaces in shear loading, thereby, reducing the local ductility. Global ductility is improved by higher mechanical phase contrast and lower martensite volume fraction, whereas local ductility is improved by low phase contrast and higher martensite volume fraction. The hardening behavior of martensite is the key to avoiding the above trade-off between local and global ductility.It is shown that if the strain hardening capacity of martensite in the later stages of deformation (in high strains) can be increased, this would result in removing the observed paradoxical trends of local and global ductility in DP steels.
    \end{abstract}
\begin{keyword}
 global ductility \sep local ductility \sep  edge cracking \sep martensite hardening \sep dual-phase steels
\end{keyword}

\end{frontmatter}


\section{Introduction}
\label{sec:introduction}

	In light of providing more detailed criteria to compare the performance of sheet metal, the terminology of 'global' and 'local' ductility/formability has been suggested in the literature \citep{Hance2016a}. 
	
	Global ductility refers to the ability of the material to undergo plastic deformation without the formation of a (localized) neck \citep{Gruenbaum2019}. As necking is the outcome of a competition between geometrical softening, and physical hardening, a higher strain hardening capacity generally improves global ductility. Due to the correlation of global ductility with the strain hardening capacity, ductility measures such as uniform elongation obtained from uniaxial tensile tests or FLC curves obtained from Nakajima type tests are relevant indicators of global ductility. Other examples of tests measuring global ductility are drawing, stretch forming, and plane-strain tension. These tests and the associated sample geometries are such that they apply a relatively uniform deformation over a comparatively large region (e.g. the gauge section with stress-free edges) before the onset of localized necking.

	The strains occurring beyond localized necking are local in nature, reflecting the local formability/ductility of the material. Accordingly, local ductility refers to the ability to undergo substantial plastic deformation in a local area without fracture  \citep{Heibel2018}. Damage tolerance or fracture resistance, which are intrinsic characteristics of a particular material, correspond well with the definition of local ductility \citep{Larour2019, Frometa2019}. The tests that characterize local ductility measure failure typically in a plane strain condition without (or with very slight) necking. Examples of such tests are notched tensile specimens, the hole expansion test, and the tight-radii bending test such as VDA 238-100 \citep{Cheong_2017, VDA}. In the uniaxial tensile test, the true fracture strain or the reduction of area (or thickness) are also regarded as representative indicators of local ductility \citep{Hance2016a, Larour2017, Heibel2018}. Local ductility correlates with the predictions of the fracture forming limit diagram (FFLD) or the shear fracture forming limit (SFFL) diagrams for sheet metal, given that these limits probe the damage tolerance in the absence of necking or in the post necking regime \citep{Marciniak2002, Isik2014}. Incremental forming techniques can stabilize/control the necking and hence measure strains occurring beyond localized necking \citep{Lu2019}. In summary, when necking is prevented, it will be the local, intrinsic ductility that limits the performance.

\begin{figure}[ht!]
\centering
\includegraphics[width=0.6\textwidth]{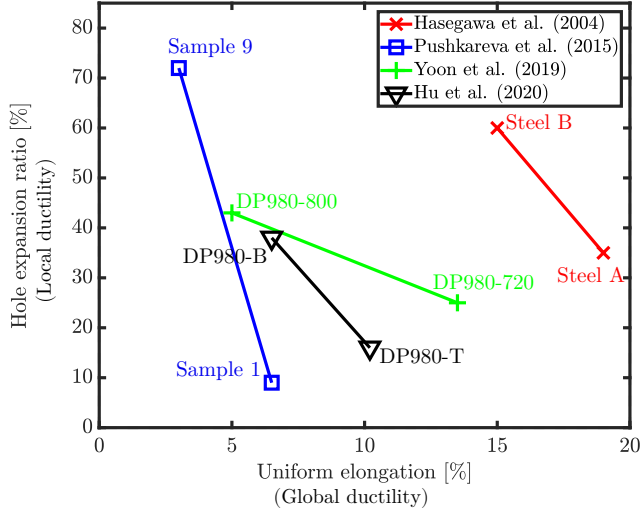}
\caption{The ductility paradox illustrated by pairs of DP steels of (nearly) the same strength and composition, subjected to tensile test and hole expansion test. Data adopted from references \citep{Hasegawa2004,Irina,Yoon2019, Hu2020}. In each case, the material with the higher ductility in the global ductility test reveals a lower ductility in the local ductility test. 
}\label{fig:GlvsLC}
\end{figure}

	In advanced high strength steels, a high level of global ductility is not necessarily accompanied by a high level of local ductility. This is most evident in the case of DP steels, where an increase in global ductility generally comes at the expense of a decrease in local ductility and vice versa. A specific example of this paradox has been identified in the sheet metal forming community for 'cut-edge failure' or 'edge cracking', i.e., the premature failure at cut-edge regions of a blank for, e.g., DP steels \citep{Davies1983, Wu2012}. This phenomenon is found to be more critical in DP grades that are classified as comparatively ductile on the basis of conventional formability tests, whereas grades with a lower global ductility tend to be more resistant against edge cracking. For instance, \citet{Hasegawa2004} measured the uniform elongation and hole expansion ratio of two DP steels of the same nominal strength, $1000$ MPa, with almost identical composition. The grade with the higher uniform elongation in the tensile test exhibited earlier edge cracking in the hole expansion test. Similar observations have been reported by \citet{Taylor2014} (cf. steels A and E), \citet{Irina} (cf. samples $1$ and $9$), \citet{Yoon2019}, and \citet{Hu2020}. In all cases, two DP steels of (nearly) the same composition and strength but different martensite volume fractions were produced and then compared in terms of particular global and local ductility measures, i.e. uniform elongation and hole expansion ratio. The data obtained in these studies is summarized in Figure \ref{fig:GlvsLC}. It consistently shows, for each pair of materials, that a higher uniform elongation entails a lower hole expansion ratio -- although the magnitude of this contrasting effect varies between the different studies. Similar observations have been reported in studies wherein the global and local ductility of DP steel is compared to that of complex phase (CP) steel \citep{Karelova2009, Heibel2018}. 

	This paper aims to rationalize the above observations and unravel the paradoxical trend governing the global and local ductility in advanced high strength steels, and particularly DP steels. We hypothesize that the paradox may be explained as follows. Since the global ductility is determined by the necking instability, to improve it the instability needs to be delayed. This can be done by introducing more hardening capacity to the DP steel. It is shown in the literature that, for a fixed composition, a lower volume fraction of martensite with an increased hardness (having a higher carbon content) leads to a higher strain hardening capacity. However, the resulting higher mechanical phase contrast (hardness difference) between ferrite and martensite also entails more straining of the ferrite, in which, as a consequence, damage is initiated earlier and grows at a higher rate. This earlier damage results in a lower local ductility. So the higher contrast between the two phases leads to a higher global ductility on the one hand, but a lower local ductility on the other hand. This paradoxical behavior would then originate from the fact that the hardening capacity of the martensite in the later stages of the deformation (in high strains) is low. Therefore, avoiding plastic deformation in it, by increasing phase contrast, should be beneficial for the global ductility and detrimental for the local ductility.  

	We test the above hypothesis by a systematic statistical analysis on a highly idealized micromechanical model of DP steel based on the approach suggested by \citet{DeGeus2017, DeGeus2016}. We construct two-dimensional, microstructural volume elements consisting of square grains of ferrite and martensite, and employ isotropic plasticity for both phases. In spite of the simplicity of these idealized microstructures, it has been shown that rich conclusions can be drawn, since they enable a systematic statistical analysis \citep{DeGeus2015}. Different DP microstructures of the same strength are produced by simultaneously varying the mechanical phase contrast and volume fractions. To produce such virtual steel grades with the same strength, the material parameters of the ferrite are fixed, whereas those of the martensite are varied by scaling its hardening curve. We further characterize the (i) hardening behavior and (ii) failure strain of the microstructures by simulating a pure shear deformation. In the numerical study, the following measures are employed to quantify the local and global ductility:
	
\begin{itemize}
\item 
    Local ductility: the point of failure under pure shear deformation, as predicted by a simple damage criterion. In pure shear, one does not expect any necking, hence it is a suitable loading condition to examine the intrinsic ductility of the generated virtual DP microstructures. 
\item
    Global ductility: the point of (localized) necking in plane strain tension (i.e. FLC point). In this case, we do expect necking to occur and the strain associated with the peak of the engineering stress-strain curve will be the indicator of global ductility. In fact, as detailed in what follows, the nominal response can be predicted easily and accurately based on the same pure shear analysis carried out to determine local ductility, and hence, only one pure shear simulation is needed to predict the global and local ductility of a microstructure.
\end{itemize}	
 
	The simulations are carried out in a fast \textsc{Fourier} transform (FFT) based spectral solver \citep{Roters2019}. It is demonstrated that, as hypothesized, increasing the mechanical contrast by reducing the martensite volume fraction results in a better global ductility, but a reduced local ductility -- as observed in the experiments. We furthermore show that, the overall hardening behavior of the martensite is the key to breaking this paradoxical behavior. 

\section{Modeling framework}
\label{sec:methodology}

\subsection{Microstructural model}
\label{sec:Microstructural model}
\begin{figure}[ht!]
\centering
\includegraphics[width=1\textwidth]{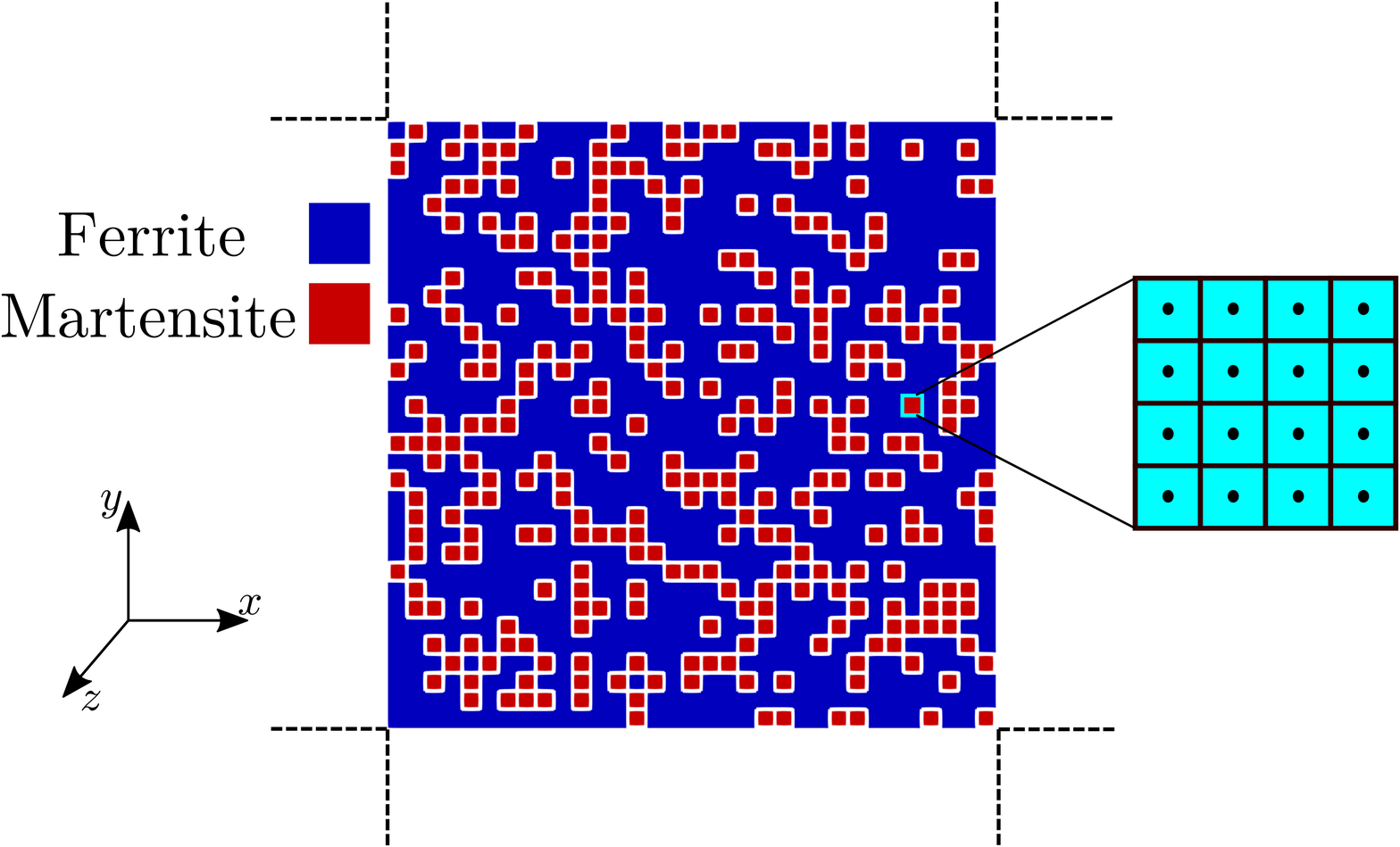}
\caption{A typical two-phase microstructural volume element of $33\times33$ grains sampled from the ensemble of $100$ random periodic volume elements. The martensite phase volume fraction is $ \varphi= 0.27$. Each grain is discretized spatially by $4\times 4$ Fourier grid points, as shown by the highlighted grain on the right.}
\label{fig:DP-micro}
\end{figure}

	The microstructural DP models which we employ consist of randomly distributed equi-sized square grains of ferrite and martensite. This is a highly idealized approach which has proven to be useful in the past to better understand the interplay between microstructure and damage processes \citep{DeGeus2015, DeGeus2016}. Its geometrical simplicity and the comparatively low computational cost associated with it allow one to consider many random realizations, enabling a thorough statistical analysis leading to relevant conclusions on the underlying mechanisms. An illustration of a $2D$ microstructural volume element (MVE) with $33\times33\approx 1000$ grains is presented in Figure \ref{fig:DP-micro}. Here, each square is representative of one ferrite or martensite grain. The reference example shown in Figure \ref{fig:DP-micro} consists of $27\%$ of martensite grains (red squares) which are distributed randomly in the microstructure. For each of the 6 combinations of hardness contrast and volume fraction $100$ realizations are generated. Periodic boundary conditions are employed to minimize boundary effects. An efficient fast \textsc{Fourier} transform (FFT) based spectral solver is employed \citep{Roters2019}. Spectral solvers use a regular grid of material points. Here, all grains are composed of $4\times4$ grid points. The simulation results are represented in terms of grain-averaged quantities to eliminate the effect of artificial stress concentrations at corners of the square cells and to reduce the effect of the \textsc{Gibbs} phenomenon \citep{Zeman2017, Wanguemert-Perez2007}. The model has been implemented in the DAMASK software \citep{Roters2019}, providing the solution of the periodic mechanical boundary value problem and hence the mechanical response of the microstructures. For more information on the framework the reader is referred to \citep{DeGeus2015,DeGeus2016}.

\subsection{Constitutive model}
\label{Modelsubsec2}
	The significant mismatch in mechanical properties between the two phases may cause high local strains and necessitates a large deformation framework. To reduce the computational cost associated with the statistical-micromechanical analysis, isotropic visco-plasticity is considered for the behavior of both phases. The martensite substructure and the significant anisotropy which it may introduce are not considered here \citep{Du2019, Maresca2016}. In spite of the these simplifications, relevant conclusions on the statistical behavior of these microstructures still emerge, as shown below. 
	
	The adopted isotropic visco-plasticity model is based on the phenomenological crystal plasticity model of reference \citep{Peirce1983}, in which the deformation gradient is decomposed multiplicatively as,
\begin{align}
\mathbf{F}=\mathbf{F}_\mathrm{e}\cdot \mathbf{F}_\mathrm{p}.
\end{align}

	The elastic part of the response is expressed in $\mathbf{F}_{\mathrm{e}}$, via the generalized \textsc{Hooke}'s law,
\begin{align}
\mathbf{S}_\mathrm{e}= \, ^4\mathbb{C}:\mathbf{E}_\mathrm{e},
\end{align}
where $^4\mathbb{C}$ denotes the 4th-order isotropic elasticity tensor, and $\mathbf{E}_e=1/2(\mathbf{C}_\mathrm{e}-\mathbf{I})$ is the elastic \textsc{Green-Lagrange} strain tensor, in which $\mathbf{C}_\mathrm{e}= \mathbf{F}_{\mathrm{e}}^{T}\cdot\mathbf{F}_{\mathrm{e}}$ is the elastic right \textsc{Cauchy-Green} deformation tensor. Note that the stress tensor $\mathbf{S}_\mathrm{e}$ lives in the intermediate (plastic) configuration; it is obtained by the push forward of the second \textsc{Piola-Kirchhoff} stress $\mathbf{S}$ as,
\begin{align}\label{eq:3}
\mathbf{S}_\mathrm{e}=\mathbf{F}_\mathrm{p}\cdot \mathbf{S} \cdot \mathbf{F}_\mathrm{p}^{T}.
\end{align}
	
	This stress measure in the intermediate configuration will later be used to drive the plasticity. The plastic part of the model is formulated in rate form. To define it, we first decompose the velocity gradient tensor $\mathbf{L}= \dot{\mathbf{F}} \cdot \mathbf{F}^{-1}$ as, 
\begin{align}\label{eq:4}
\mathbf{L}= \mathbf{L}_{\mathrm{e}} + \mathbf{F}_\mathrm{\mathrm{e}} \cdot \mathbf{L}_{\mathrm{\mathrm{p}}}\cdot \mathbf{F}_{\mathrm{e}}^{-1},
\end{align}
where $\mathbf{L}_{\mathrm{e}}= \dot{\mathbf{F}}_{\mathrm{e}} \cdot \mathbf{F}_{\mathrm{e}}^{-1}$ is the elastic velocity gradient. The plastic velocity gradient $\mathbf{L}_\mathrm{\mathrm{p}}$ is assumed to be colinear with the deviatoric stress,
\begin{equation}
\mathbf{L}_{\mathrm{p}}= \dot{\mathbf{F}}_{\mathrm{p}} \cdot \mathbf{F}_{\mathrm{p}}^{-1}=\dfrac{\dot{\gamma}_{\mathrm{p} }}{T} \dfrac{\mathbf{M}_{\mathrm{p}}^{\mathrm{dev}}}{ \parallel  \mathbf{M}_{\mathrm{p}}^{\mathrm{dev}} \parallel_{F}}.   
\end{equation}
Here, $\dot{\gamma}_{\mathrm{p}}$ represents the slip rate on an (imaginary) slip system, the (\textsc{Taylor}) factor $T$ relates it to the equivalent strain rate $\dot{\bar{\varepsilon}}_{\mathrm{p}}$, $\mathbf{M}_{\mathrm{p}}^{\mathrm{dev}}$ is the deviatoric part of the \textsc{Mandel} stress tensor, $\mathbf{M}=\mathbf{C}_{\mathrm{e}} \cdot \mathbf{S}_{\mathrm{e}}$ (which is in the same configuration as $\mathbf{S}_{\mathrm{e}}$), and $\parallel \cdot \parallel_{F}$ is the \textsc{Frobenius} norm. The slip rate $\dot{\gamma}_{\mathrm{p}}$ is formulated as a power-law kinetic equation \citep{Hutchinson1976}, which is driven by $J_2$, i.e. the second invariant of the deviatoric \textsc{Mandel} stress,

\begin{equation}
\dot{\gamma}_\mathrm{p}=\dot{\gamma}_0\Bigg( \dfrac{\sqrt{ 3J_{2}}}{T \xi} \Bigg)^{n}=\dot{\gamma}_0\Bigg(\sqrt{\dfrac{3}{2}}\dfrac{\parallel  \mathbf{M}_{\mathrm{p}}^{\mathrm{dev}} \parallel_{F}}{T\xi}    \Bigg)^{n},
\end{equation}
where the constant $\dot{\gamma}_0$ is the reference shear rate and the exponent $n$ is the strain rate sensitivity. The ratio $\sqrt{3J_{2}}/T$ represents a characteristic resolved shear stress. It is compared against (divided by) the slip resistance $\xi$ which acts as the scalar internal state variable that captures the resistance to plastic deformation. $\xi$ evolves from its initial value, $\xi_{0}$, towards a saturation value, $\xi_{\infty}$, according to the evolution law \citep{Brown1989, Bronkhorst1992},
\begin{align}\label{flowressistane}
\dot{\xi}= \dot{\gamma}_{\mathrm{p}} \, h_0 \, \left| 1- \dfrac{\xi}{\xi_\infty}\right|^{a}\, \mathrm{sgn}\left( 1- \dfrac{\xi}{\xi_\infty}\right),
\end{align}
in which, $a$ is a hardening shape factor, and $h_{0}$ is a hardening modulus.

\subsection{Constitutive parameters}\label{Sec:parameters}
\begin{figure}[ht!]
\centering
\includegraphics[width=0.7\textwidth]{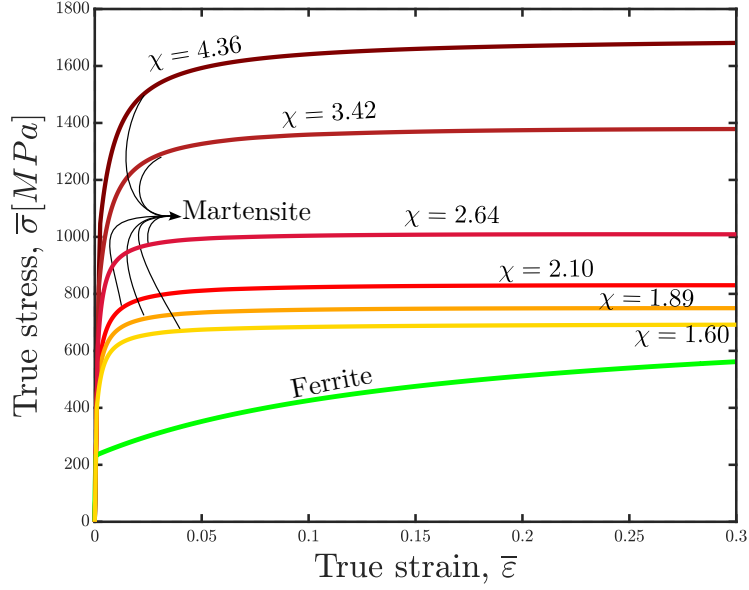}
\caption{True (equivalent) stress-strain response of the constituent phases. The ferrite behavior is the same for all grades, whereas the martensite behavior varies with the value of the mechanical phase contrast $\chi$. The variation of the martensite behavior changes in accordance with the overall martensite volume fraction, which is inversely proportional the average carbon concentration in the martensite islands. } 
\label{fig:PhaseBehaviors}
\end{figure}

	The set of parameters for ferrite and martensite is adopted from \citep{Tasan2014c,Naunheim2020} and is given in Table \ref{Ch3--tab:PhaseBeh}. The \textsc{Taylor} factor is taken from the literature, where it is obtained from CPFEM analysis for BCC crystals with one slip family $\{110\}$ \citep{Rosenberg1971,Zhang2019}. The mechanical contrast between the phases is manipulated by scaling the initial flow stress of the martensite up and down using the mechanical phase contrast parameter, $\chi$. The martensite hardening response obtained for several values of $\chi$ is shown in Figure \ref{fig:PhaseBehaviors}, together with that of ferrite. This alteration of the martensite behavior represents the change of carbon content due to a change in volume fraction of martensite, whereas the ferrite behavior is fixed for all volume fractions due to the low solubility of the carbon \citep{Chen1989,Krauss1999a}. Literature has reported that an increase in carbon content of the martensite increases the associated strain hardening \citep{allain2012toward,Krauss1999a}. This effect is qualitatively taken into account here, see Figure \ref{fig:PhaseBehaviors}. It is assumed that the carbon concentration is the same in all martensite islands. The critical points to consider in the behavior of the martensite are to incorporate a high initial hardening, of the same order of magnitude as the \textsc{Young}'s modulus, accompanied with a fast saturation. These features of martensite behavior have been frequently reported \citep{zaccone1993elastic,Tasan2014c,hutchinson2018plastic,Naunheim2020}. 
\begin{table}[ht!]
  \caption{Parameter values of the constituent phases used in the simulations}
  \centering
  \begin{threeparttable}
    \begin{tabular}{cc@{\qquad}cc}
      Property & Unit & Ferrite & Martensite \\ \midrule\midrule 
      
        \makecell{$K$} & $GPa$ & $176$ & $314$  \\      
        \makecell{$G$} & $GPa$ & $82.5$ & $147$  \\       \cmidrule(l r){1-4}
        \makecell{$\dot{\gamma}_{0}$} & $s^{-1}$ & $0.001$ & $0.001$ \\      
        \makecell{$\xi_{0}$} & MPa & $95.5$ & $\chi \times \xi_{0}^{F}$  \\
        \makecell{$\xi_{\infty}$} & MPa & $317.35$ & $1.7 \times \xi_{0}^{M}$  \\ 
        \makecell{$h_{0}$} & $GPa$ & $1$ & $563$  \\      
        \makecell{$n$} & -- & $20$ & $20$  \\      
        \makecell{$a$} & -- & $2.25$ & $2.25$  \\    
        \makecell{$T$} & -- & $2.4$ & $2.4$  \\ 
        \cmidrule(l r){1-4}  
        \makecell{$\chi$ } & -- & $--$ & $1.66-4.36$  \\
        	\cmidrule(l r){1-4}  
        \makecell{$A$} & -- & $0.2$ & $0.87$  \\
        \makecell{$B$} & -- & $1.7$ & $5.6$  \\
        \makecell{$\varepsilon_{\mathrm{PC}}$} & -- & $0.1$ & $0$  \\
        \midrule\midrule
    \end{tabular}
  \end{threeparttable}
  \label{Ch3--tab:PhaseBeh}
  \end{table}

\subsection{Loading condition}
\label{Sec:Loading}
	As characteristic to the spectral method, the boundary conditions of the MVE shown in Figure \ref{fig:DP-micro} are periodic. The MVE is subjected to a mean plane strain pure shear (PS) loading along the main axes, where the applied macroscopic velocity gradient tensor reads,
\begin{equation}\label{eq:loading}
\langle\mathbf{L}\rangle=\langle\mathbf{D}\rangle=\dot{\varepsilon} (\vec{e}_{x}\vec{e}_{x}-\vec{e}_{y}\vec{e}_{y}),
\end{equation}
with $\dot{\varepsilon}=0.001\, s^{-1}$ the applied logarithmic strain rate, $\mathbf{D}$ symmetric rate of deformation tensor, and $\langle\rangle$ means the quantity within is averaged over the MVE and represented in the macro-scale. This choice of loading condition prevents necking in the microstructure.  

\section{Damage criteria -- local ductility}
\label{Sec:Local}
	In the numerical study, the local ductility is quantified through the damage tolerance of a particular microstructure under pure shear deformation, as described in Section \ref{Sec:Loading}. The damage is predicted based on the accumulated plastic strain $\varepsilon_{\mathrm{p}}$ and stress triaxiality $\eta=M^\mathrm{hyd}/\bar{M}$, which is the ratio of the hydrostatic stress to the equivalent \textsc{von Mises} stress in terms of the \textsc{Mandel} stress. Here, the \textsc{Johnson-Cook} (JC) \citep{Johnson1985} damage model is used in which the effective plastic strain rate $\dot{\varepsilon}_{\mathrm{p}}=\sqrt{\dfrac{2}{3}\mathbf{D}_{\mathrm{p}} \cddot \mathbf{D}_{\mathrm{p}}}$ is incrementally compared to a critical strain $\varepsilon_{\mathrm{c}}(\eta)$, that depends on the stress triaxiality,
\begin{equation} \label{eq:damage}
D= \int_{0}^{t} \dfrac{\dot{\varepsilon}_\mathrm{p}}{\varepsilon_{\mathrm{c}}(\eta)}\mathrm{d}t \qquad \mathrm{with} \qquad \varepsilon_{\mathrm{c}}= A\mathrm{exp}(-B\eta)+\varepsilon_{\mathrm{pc}}
\end{equation}
wherein $A$, $B$, and $\varepsilon_{\mathrm{pc}}$, are material parameters \citep{Vajragupta2012a,Bareggi2012}; and their values are listed in Table \ref{Ch3--tab:PhaseBeh}.
	
	At the end of each time step of the simulation the increment of damage $D$ in a particular realization is averaged over all grain integration points and then accumulated through the loading history. Initially $D=0$, while $D=1$ indicates failure. Here, the damage is only used as an indicator -- i.e. it does not affect the mechanics. As soon as $1 \%$ of all the grains in the microstructure fail, i.e. $D\geqslant1$, global failure is assumed to occur and the corresponding overall fracture strain is extracted from the volume element. The fracture strains are then averaged over all the realizations. This failure criteria, and particularly the requirement that $1\%$ of the grains fails, is representative for dual-phase steels, see, e.g., \citep{Tasan2009a}, while it has been verified that the trends are not sensitive to it \citep{DeGeus2016b, DeGeus2017}. The applied load case is globally volume-preserving, see Section \ref{Sec:Loading}, and hence any non-zero triaxiality arising in the microstructure can only be attributed to the interactions of the phases with different properties.

\section{Necking criteria --  global ductility} \label{Necking-global}
	To quantify the global ductility a conventional criterion based on the onset of localized necking (plastic instability) is adopted. This macro-scale necking-controlled failure criterion signals the loss of load bearing capacity in the steel based on the onset of a tensile instability, i.e. the point where the engineering stress drops.
 
	To predict the peak of the nominal stress-strain curve of a certain virtual steel grade, we first extract the respective true stress versus true strain response (both equivalent) from the pure shear analysis explained in Section \ref{Sec:Loading}. Next, these quantities are used to predict the engineering stress-strain response of the same virtual grade under a plane strain tension test. Finally, the onset of localized necking and thus global ductility is determined as the maximum engineering stress. The details and related assumptions to predict the engineering response of a plane strain tension test from a pure shear analysis are described below. 

	By incorporating Eqn. \ref{eq:loading} and assuming that the elastic strains are negligible (rigid plasticity), the rate of macroscopic effective strain is,
\begin{align}\label{eq:PS3}
&\langle\dot{\bar{\varepsilon}}\rangle=\sqrt{\dfrac{2}{3} \langle\mathbf{D} \rangle: \langle\mathbf{D}\rangle}= \dfrac{2}{\sqrt{3}} \dot{\varepsilon},  
\end{align}
and thus the macroscopic equivalent (plastic) strain equals, $\langle\bar{\varepsilon}\rangle=\dfrac{2}{\sqrt{3}} \varepsilon$. The rate of deformation tensor, i.e. the flow rule, in large strain theory is defined as, $\langle\mathbf{D}\rangle= \dot{\bar{\varepsilon}} \dfrac{3}{2} \dfrac{\langle\sigma^{\mathrm{dev}}\rangle}{\langle\bar{\sigma}\rangle}$. Therefore, by incorporating Eqn. \ref{eq:loading} and \ref{eq:PS3}, the macroscopic deviatoric stress tensor $\langle\sigma^{\mathrm{dev}}\rangle$ reads,
\begin{align}\label{eq:PS4}
&\langle\mathbf{\boldsymbol{\sigma}^{\mathrm{dev}}}\rangle= \dfrac{2}{3} \langle\bar{\sigma}\rangle \dfrac{\langle\mathbf{D}\rangle}{\langle\dot{\bar{\varepsilon}}\rangle}= \dfrac{\langle\bar{\sigma}\rangle}{\sqrt{3}} (\vec{e}_{x}\vec{e}_{x} -\vec{e}_{y}\vec{e}_{y}).
\end{align}
In plane strain tension (PT) conditions, the following mixed boundary conditions are enforced on average in macro-scale,
\begin{align}\label{eq:PSload}
& \langle\mathbf{F}_{\mathrm{PT}}\rangle=\begin{bmatrix}
\mathrm{exp}(\varepsilon) & 0 & 0 \\
0 & \ast & 0 \\
0 & 0 & 1
\end{bmatrix}, \, \, \langle\mathbf{\mathbf{\mathbf{P}}}_{\mathrm{PT}}\rangle=\begin{bmatrix}
 \ast & \ast & \ast\\
\ast & 0 & \ast \\
\ast & \ast& \ast
\end{bmatrix}, 
\end{align}
where $\mathbf{P}$ denotes the first \textsc{Piola-Kirchhoff} stress tensor, which is the work conjugated stress to the deformation gradient tensor, $\mathbf{F}$. Here, $\ast$ indicates that a particular component of $\mathbf{F}$ or $\mathbf{P}$ is free to evolve, and hence follows from the simulation. Assuming rigid-plastic behavior, the deformation gradient tensors $\mathbf{F}$ for the two situations, i.e. pure shear and plane strain tension, have to be identical, and the only difference is a hydrostatic contribution to the stress tensor for the plane strain tension case (PT) which is absent for the pure shear case. As shown in Eqn. \ref{eq:PSload}, $\langle P_{yy} \rangle=0$ and considering the relation $\mathbf{\boldsymbol{\sigma}}=J^{-1}\mathbf{P} \cdot \mathbf{F}^{\mathrm{T}}$ we hence have,
\begin{align}\label{eq:PSSS5}
& \langle\mathbf{\sigma}_{yy}\rangle= \langle-p \rangle-\dfrac{\langle\bar{\sigma}\rangle}{\sqrt{3}}=0,
\end{align}
where $p$ is the hydrostatic pressure, which shows that:
\begin{align}\label{eq:PS5}
\langle-p\rangle= \dfrac{\langle\bar{\sigma}\rangle}{\sqrt{3}}. 
\end{align}
By taking Eqn. \ref{eq:PS5} into account, the relation between the stress tensor of the plane strain tension and equivalent stress reads,
\begin{align}\label{eq:PS6}
&\langle\boldsymbol{\sigma}\rangle= \dfrac{\langle\bar{\sigma}\rangle}{\sqrt{3}} (2\vec{e}_{x}\vec{e}_{x} +\vec{e}_{z}\vec{e}_{z}).
\end{align}
As a result, for the case of plane strain tension $\langle\sigma_{xx}\rangle= \dfrac{2}{\sqrt{3}}\langle\bar{\sigma}\rangle$. Let us consider the stress and strain transformation between the reference and current configuration, $\mathbf{P}= J\boldsymbol{\sigma} \cdot \mathbf{F}^{-\mathrm{T}}$, and the \textsc{Green-Lagrange} strain $\mathbf{E}$. Then, the engineering stress-strain response of plane strain pure shear analysis is given in terms of macroscopic equivalent stress, $\langle\bar{\sigma}\rangle$, and equivalent strain, $\langle\bar{\varepsilon}\rangle$, by,
\begin{align}\label{eq:PS7}
\langle P_{xx}\rangle= \dfrac{2}{\sqrt{3}}\langle \bar{\sigma}\rangle \, \, \mathrm{exp}\left(-\dfrac{\sqrt{3}}{2}\langle \bar{\varepsilon} \rangle\right) \qquad \mathrm{and} \qquad \langle E_{xx}\rangle= \mathrm{exp}\left(\dfrac{\sqrt{3}}{2}\langle\bar{\varepsilon}\rangle\right)-1.
\end{align} 
The global ductility is determined by the value of the engineering strain, $\langle E_{xx}\rangle$, in which the engineering stress, $\langle P_{xx}\rangle$, is maximum. The main advantage of this method is that both local and global ductility can be predicted from the same simulation.

\section{Results \& Discussion}
\label{Sec:Results}

\subsection{Virtual steel grades of similar strength}
\begin{figure}[ht!]
\centering
\includegraphics[width=0.9\textwidth]{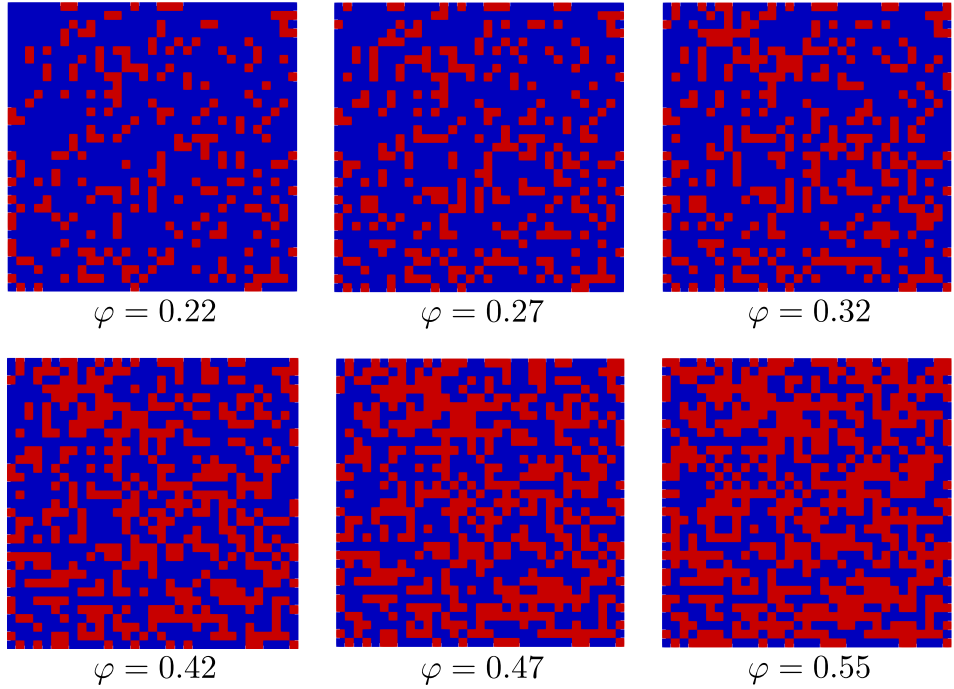}
\caption{Two-phase microstructural volume elements sampled from each ensemble of generated virtual $DP600$ grades.
 The martensite volume fraction is varied within the range of  $\varphi=0.22$ to $\varphi=0.55$.}
\label{fig:geoms}
\end{figure}

\begin{figure}[ht!]
\centering
\includegraphics[width=0.57\textwidth]{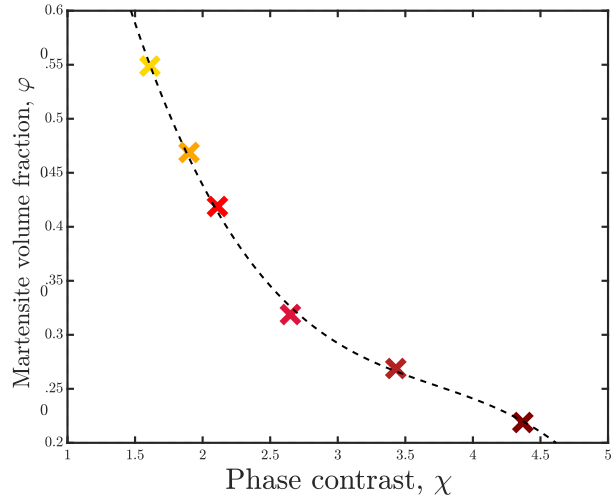}
\caption{Combinations of values for $\varphi$ and $\chi$ used to generate ensembles of virtual $DP600$ steel grades. The dotted line is a nonlinear trend line. To reach the same strength, an increase in $\varphi$ necessitates a decrease in $\chi$. } 
\label{fig:ContvsVoloum}
\end{figure}

	Ensembles of virtual DP steel microstructures with similar ultimate strength were generated manually by tuning the volume fraction and strength of the martensite phase. 	Figure \ref{fig:geoms} shows one realization chosen (randomly) from each ensemble of 100 realization. The ensembles differ only in their martensite volume fraction $\varphi$, and mechanical phase contrast $\chi$; all other material and microstructural parameters are identical. The combinations of $\varphi$ and $\chi$ used to generate the virtual grades are shown in Figure \ref{fig:ContvsVoloum}. An increase in the martensite volume fraction must be accompanied by a decrease in the mechanical phase contrast to provide a constant strength. This combined variation reflects the effect of the carbon content distribution on the average martensite behavior \citep{Lubahn1968, Krauss1999a, morsdorf2021carbon}.

\begin{figure}[ht!]
\centering
\includegraphics[width=1\textwidth]{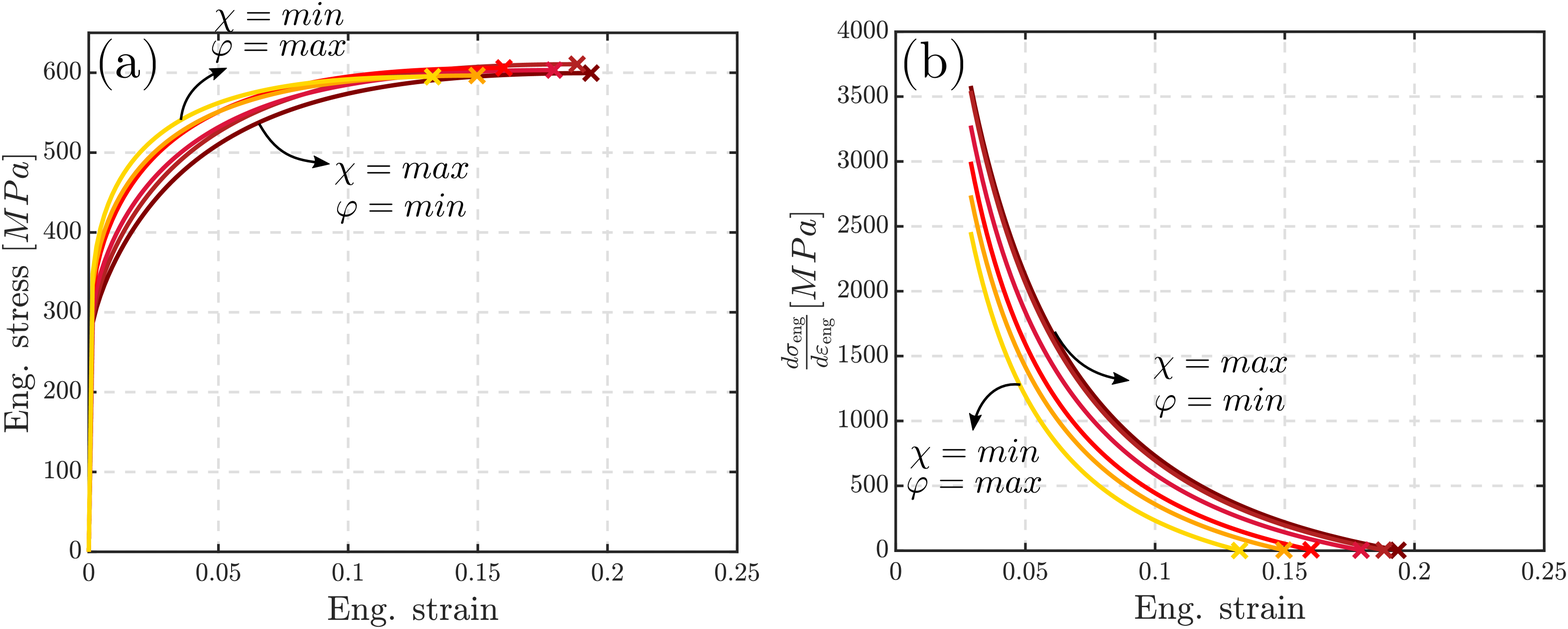}
\caption{a) The engineering response of the generated virtual steel grades in plane strain tension as predicted based on the computed pure shear response. b) The strain hardening rate of these grades extracted from the engineering responses. In both plots cross markers $(\times)$ indicate the onset of necking, $d\sigma_{\mathrm{eng}}/d\varepsilon_{\mathrm{eng}}=0$, which correspond to the uniform elongation of the steels. }
\label{fig:GL_AF}
\end{figure}

	The ensemble-averaged engineering stress-strain curves (for plane strain tension) of the generated grades are shown in Figure \ref{fig:GL_AF}a. The marked peaks of the nominal curves indicate the onset of necking, and correspond to the nominal strength and uniform elongation strain. The curves have been truncated at this point, because beyond it the way they have been determined, as detailed in Section \ref{Necking-global}, is no longer valid due to localisation. Figure \ref{fig:GL_AF}b shows the strain hardening rate of the same virtual steel grades. All 6 virtual grades have been tuned, by changing $\chi$ and $\varphi$, to have an approximate nominal strength of $600$ MPa (within a maximum deviation of $10$ MPa).

\subsection{Global ductility}
	
    The uniform elongation and hence global ductility of the virtual steels grades shown in Figure \ref{fig:GL_AF} have a significant variation. The grade with the lowest martensite volume fraction, $\varphi=0.22$, and the highest mechanical phase contrast, $\chi=4.36$, reveals the highest hardening rate (see Figure \ref{fig:GL_AF}b), and hence the largest uniform elongation, $\approx 0.19$ . The lowest hardening rate and uniform elongation, $\approx 0.13$, occur for the grade with the highest martensite volume fraction, $\varphi=0.55$ and lowest phase contrast, $\chi=1.60$. Therefore, global ductility, in total, goes down roughly by the factor of one-third.

\begin{figure}[ht!]
\centering
\includegraphics[width=1\textwidth]{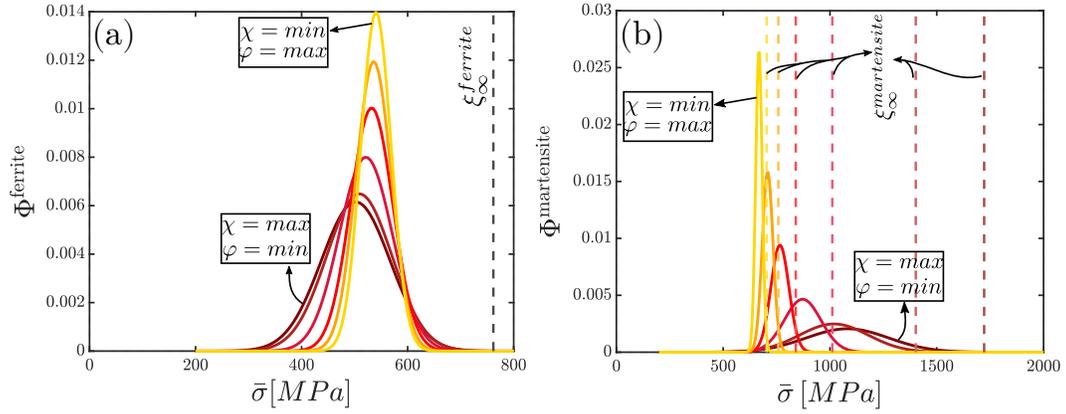}
\caption{The probability density $\Phi$ of the equivalent stress $\bar{\sigma}$ distribution for the a) ferrite, and b) martensite grains in the analysed virtual steel ensembles. The results were obtained at a global applied strain of $\bar{\varepsilon} \approx 0.2$ under pure shear deformation. The dashed lines indicate the saturation value of the phases in each grade. }
\label{fig:Strss_hist}
\end{figure}

	For each ensemble, the equivalent stress probability density distribution plot of all the ferrite and martensite grains is shown in Figure \ref{fig:Strss_hist}. The curves are plotted at a global applied strain of $\langle\bar{\varepsilon} \rangle \approx 0.2$ under pure shear deformation. The dashed lines represent the saturation stress value for each phase in the different grades. The material parameters of the ferrite are fixed, whereas the phase contrast value, $\chi$, scales the martensite properties. Figure \ref{fig:Strss_hist}a shows that the average value of the stress in the ferrite grains does not reveal significant changes, even though the heterogeneity of the distribution increases, as the phase contrast increases. The mean stress and the heterogeneity of distributions vary significantly in the martensite grains for the different grades. It is observed in Figure \ref{fig:GL_AF}b that DP grades with less but harder martensite have a higher strain hardening capacity. As shown in Figure \ref{fig:Strss_hist}b, in such grades (dark colors) a comparatively lower amount of grains have reached to their respective saturation stress; For instance, the mean stress of grains in the grade with the highest phase contrast, $\chi=4.36$, is approximately $1200$ MPa and saturation stress is $\xi_{\infty}=1750$ MPa.	Based on these observations, a higher global ductility emerges for a DP grade with a high mechanical phase contrast. 

\subsection{Local ductility}

\begin{figure}[ht!]
\centering
\includegraphics[width=0.6\textwidth]{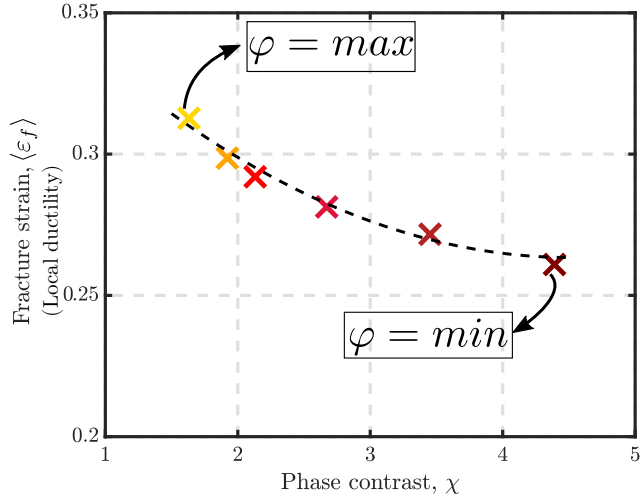}
\caption{For the generated virtual steel grades, average macroscopic fracture initiation strain, i.e., local ductility, as a function of phase contrast $\chi$ is shown. The dashed curve indicates the trend of decreasing local ductility by increasing mechanical phase contrast. } 
\label{fig:LocalDuctility}
\end{figure}

	Next, the ensemble-averaged fracture strain of each grade, determined by the damage criteria outlined in Section \ref{Sec:Local}, is compared in Figure \ref{fig:LocalDuctility}. It is shown that the best local ductility performance, $\approx 0.31$, corresponds to the grade with the highest martensite volume fraction, $\varphi=0.55$, and the lowest mechanical phase contrast, $\chi=1.60$. The grades with a high mechanical phase contrast (and a low martensite volume fraction) show a lower fracture strain, and thus a poor local ductility. Compared to the minimum phase contrast, $\chi=1.60$, the local ductility is dropped by $\approx 17\%$ for the grade with maximum phase contrast, $\chi=4.36$ .

\begin{figure}[ht!]
\centering
\includegraphics[width=1\textwidth]{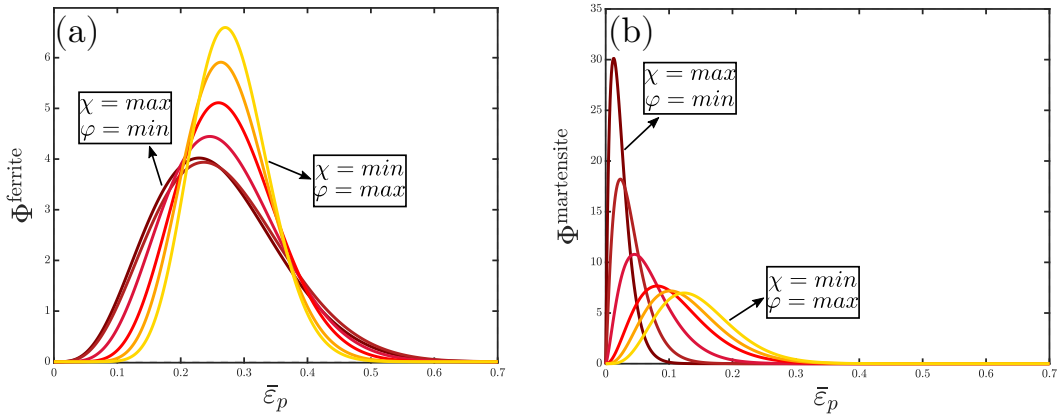}
\caption{The probability density $\Phi$ of the equivalent plastic strain $\bar{\varepsilon}_p$ distribution for the a) ferrite, and b) martensite grains in the analysed virtual steel ensembles. The results were obtained at a global applied strain of $\langle\bar{\varepsilon} \rangle \approx 0.2$. Due to the different strain distributions within both phases, the vertical axis is different in the two diagrams.}
\label{fig:Strain_hist}
\end{figure}

	Figure \ref{fig:Strain_hist} compares the strain distributions of both phases for the different virtual grades shown in Figure \ref{fig:GL_AF}. The presented results are obtained at a global applied strain of $\langle\bar{\varepsilon} \rangle \approx 0.2$ under a pure shear deformation. Note again that the material parameters of the ferrite phase are the same for all the grades. The strain distribution in the ferrite grains with a higher $\chi$ (and hence a lower $\varphi$) is highly heterogeneous (i.e. a high variance in the distribution). The hardness mismatch between the phases results in localization of deformation, predominantly in the ferrite grains. As shown in Figure \ref{fig:Strain_hist}a, the ferrite grains have to accommodate local strain levels up to $0.7$ for the applied global strain of $0.2$. These high local strain levels promote early damage of the microstructure. For lower mechanical phase contrast, the contribution of the martensite to the plasticity is higher, see Figure \ref{fig:Strain_hist}b, which leads to a more uniform deformation in the microstructure. However, as already observed in Figure \ref{fig:GL_AF}, grades with high volume fractions of martensite tend to neck earlier, revealing a poor global ductility. 

\subsection{Comparison of global and local ductility}\label{SubSec:GLvsLC}

\begin{figure}[ht!]
\centering
\includegraphics[width=0.6\textwidth]{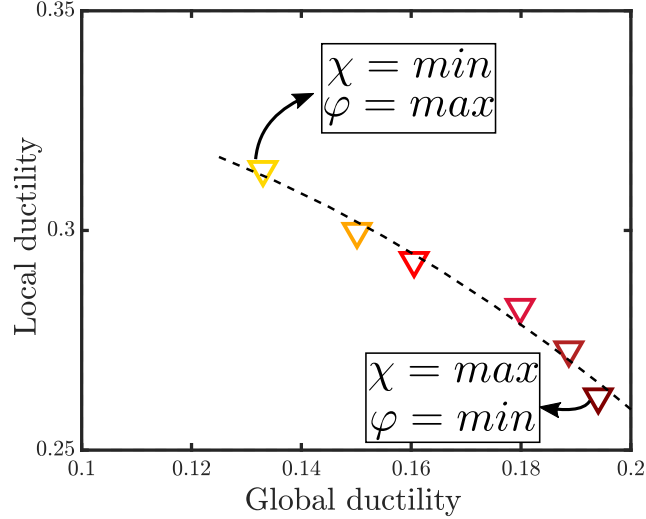}
\caption{Relation between the global and local ductility of the six virtual steel grades. The dashed curve indicates the trend of decreasing local ductility with increasing global ductility.}
\label{fig:AFSteel_GL}
\end{figure}

	Figure \ref{fig:AFSteel_GL} compares the resulting trends in terms of global and local ductility for the grades considered. As global ductility improves by, relatively speaking, $\approx 50 \%$, the local ductility decreases by $\approx 17\%$. This is in agreement with the literature, where uniform elongation and HER of DP steels (with similar strength and composition) have been compared \citep{Irina, Yoon2019,Hu2020}. Therefore, the paradox or the opposite trends observed in global and local ductility has been captured/reproduced here using a simple micromechanical model. It is shown that this trend (paradox) may be explained by the fact that less but harder martensite results in (i) stronger hardening and hence better global ductility, but also (ii) faster damage growth and hence poorer local ductility.    
	  
\begin{figure}[ht!]
\centering
\includegraphics[width=1\textwidth]{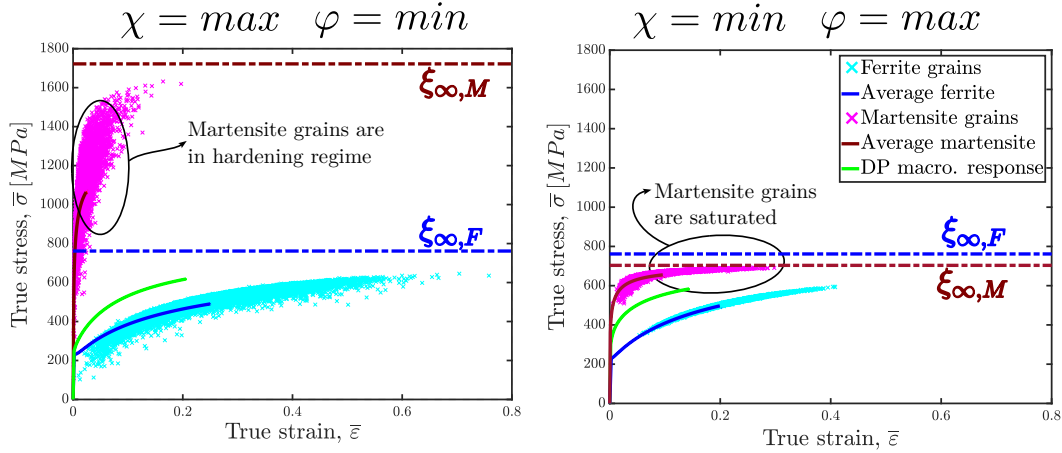}
\caption{True stress-strain response of the two virtual grades with the lowest and the highest mechanical phase contrast and martensite volume fraction. The behavior of all ferrite and martensite grains, at the instant of reaching the global ductility limit (uniform elongation) are aggregated from $20$ realizations and highlighted by cyan and magenta markers, respectively. The average behavior of the phases is illustrated with thick lines superimposed on the scatter plot of the marked individual grains}
\label{fig:Grains}
\end{figure}

	To better understand the mechanisms involved in the observed paradox between global and local ductility, the behavior of the individual grains for the two grades with the lowest and the highest mechanical phase contrast are tracked through 20 realizations and visualized for the moment that the strain corresponding to the global ductility of each steel is reached. The results are shown in Figure \ref{fig:Grains}. The average behavior of the phases is illustrated with thick lines superimposed on the scatter plot of the marked individual grains. The saturation stresses of the two phases in each case are shown by dashed lines in blue and red for ferrite and martensite, respectively. Based on the stress response of the martensite grains, indicated by magenta markers, it is shown that most of the martensite grains in the microstructure with a low phase contrast (see Figure \ref{fig:Grains}b) have saturated, i.e. the magenta markers are reached/close to the level of red dashed line, while in the microstructure with a high phase contrast, see Figure \ref{fig:Grains}a, their stress levels are still in the initial hardening range. In contrast, in the steel with a high mechanical phase contrast, Figure \ref{fig:Grains}a, ferrite grains (cyan markers) have to strain considerably more, with maximum around $0.6$, compared to the other steel, with maximum around $0.4$. 
    
    As discussed in Section \ref{sec:introduction}, an increase in the strain hardening of the material will delay the necking and hence improve global ductility. The high initial hardening (almost at the level of \textsc{Young}'s modulus) and fast saturation of martensite \citep{zaccone1993elastic, allain2012toward, hutchinson2018plastic} is unfavorable for necking-controlled failure, i.e. global ductility, since once the saturation stage is reached, the grade is unable to harden as the applied deformation further increases. In contrast, the hardening capacity of the ferrite is higher in later stages of the deformation (in higher strains) compared to that of martensite, in which it is only high at low strains, see Figure \ref{fig:PhaseBehaviors}. The high hardening capacity of ferrite combined with a high initial hardening of martensite helps steels with the high mechanical phase contrast to be more necking resistant-- see Figure \ref{fig:Grains}a. Therefore, a higher mechanical phase contrast entails a better strain hardening and hence better global ductility. As a result, in necking-controlled deformation conditions, DP microstructures should be processed such that a high amount of plasticity in martensite is avoided. This can be achieved by increasing the mechanical phase contrast between the two phases and decreasing the martensite volume fraction. 

\begin{figure}[ht!]
\centering
\includegraphics[width=1\textwidth]{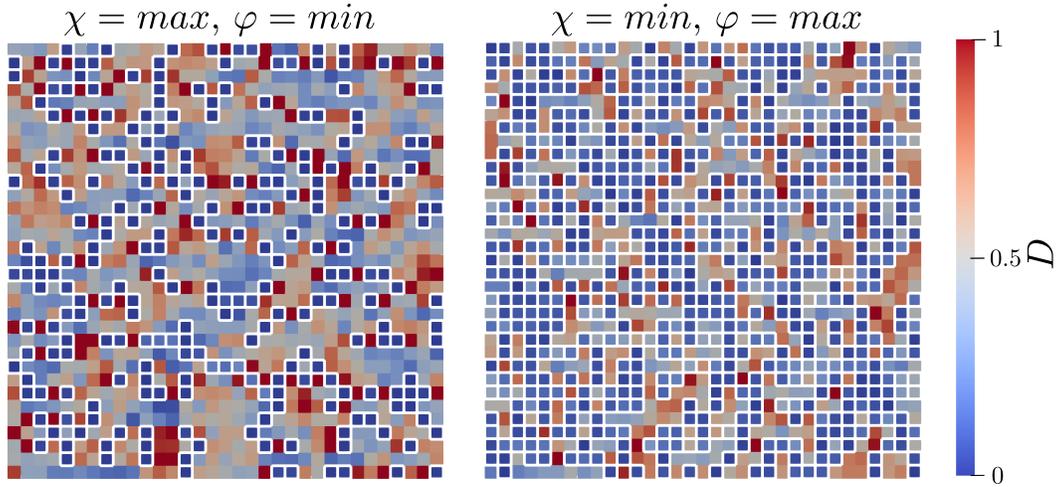}
\caption{Damage distribution map for a microstructural volume element sampled from the ensembles with extreme phase contrast and martensite volume fraction. The snapshots are taken at a global applied strain of $\langle \bar{\varepsilon}\rangle \approx 0.3$. $17\%$ of grains are damaged in the microstructure with the high phase contrast of $\chi=4.36$ and low martensite volume fraction of $\varphi=0.22$, whereas in the microstructure with the low phase contrast of $\chi=1.6$ and high martensite volume fraction of $\varphi=0.55$, only $6\%$ of grains are damaged. The martensite grains are marked by white frames. 
}
\label{fig:damage_Evolev}
\end{figure}

	Whereas avoiding plasticity in martensite by increasing the phase contrast provides a better global ductility, it also results in a poor local ductility response. The result in Figure \ref{fig:AFSteel_GL} shows that the improvement in global ductility due to a higher mechanical phase contrast compromises the local ductility. It is intuitive that a higher hardness mismatch between the two phases significantly promotes damage and failure (due to shear fracture) in the microstructure. The imposed heterogeneity results in the accommodation of high strains in the ferrite -- see Figure \ref{fig:Strain_hist}a and \ref{fig:Grains}a, leading to faster damage evolution. This is shown in Figure \ref{fig:damage_Evolev} where the damage distribution for one realization chosen from each of the two extreme ensembles is compared at a fixed applied global strain of $\bar{\varepsilon} \approx 0.3$. A higher phase contrast entails a pronounced mechanical incompatibility between the phases which results in strong stress-strain gradients all over the microstructure. By decreasing the phase contrast and increasing the martensite volume fraction, the amount of damaged grains decreases significantly. This is shown in Figure \ref{fig:damage_Evolev}, where in the microstructure with the high phase contrast $\chi=4.36$, $17\%$ of grains are damaged, i.e. $D=1$, whereas in the microstructure with the low phase contrast $\chi=1.6$, only $6\%$ of grains are damaged. A higher damage tolerance or fracture resistance of the steel is associated with a large post-uniform elongation \citep{Larour2019}, which is directly correlated with a higher HER and better edge cracking performance \citep{Frometa2019,Yoon2019}.

\subsection{Possible remedy for the ductility paradox}\label{SubSec:remedy}
	It is discussed in Section \ref{SubSec:GLvsLC} that the martensite hardening behavior can possibly be the main factor governing the paradox of local and global ductility. It is shown that the lack of strain hardening capacity at the later stages of deformation compromises global ductility. Therefore, it maybe expected that providing more (and sustained) hardening to the martensite response will improve the global ductility and hence remedy the ductility paradox. In this section, the effect of the change in the average hardening behavior of martensite is investigated. It should be emphasized that this is a theoretical exercise, in the sense that it may not be feasible to produce martensite grains with the desired properties \citep{hutchinson2011microstructures}. 
	
\begin{figure}[ht!]
\centering
\includegraphics[width=0.6\textwidth]{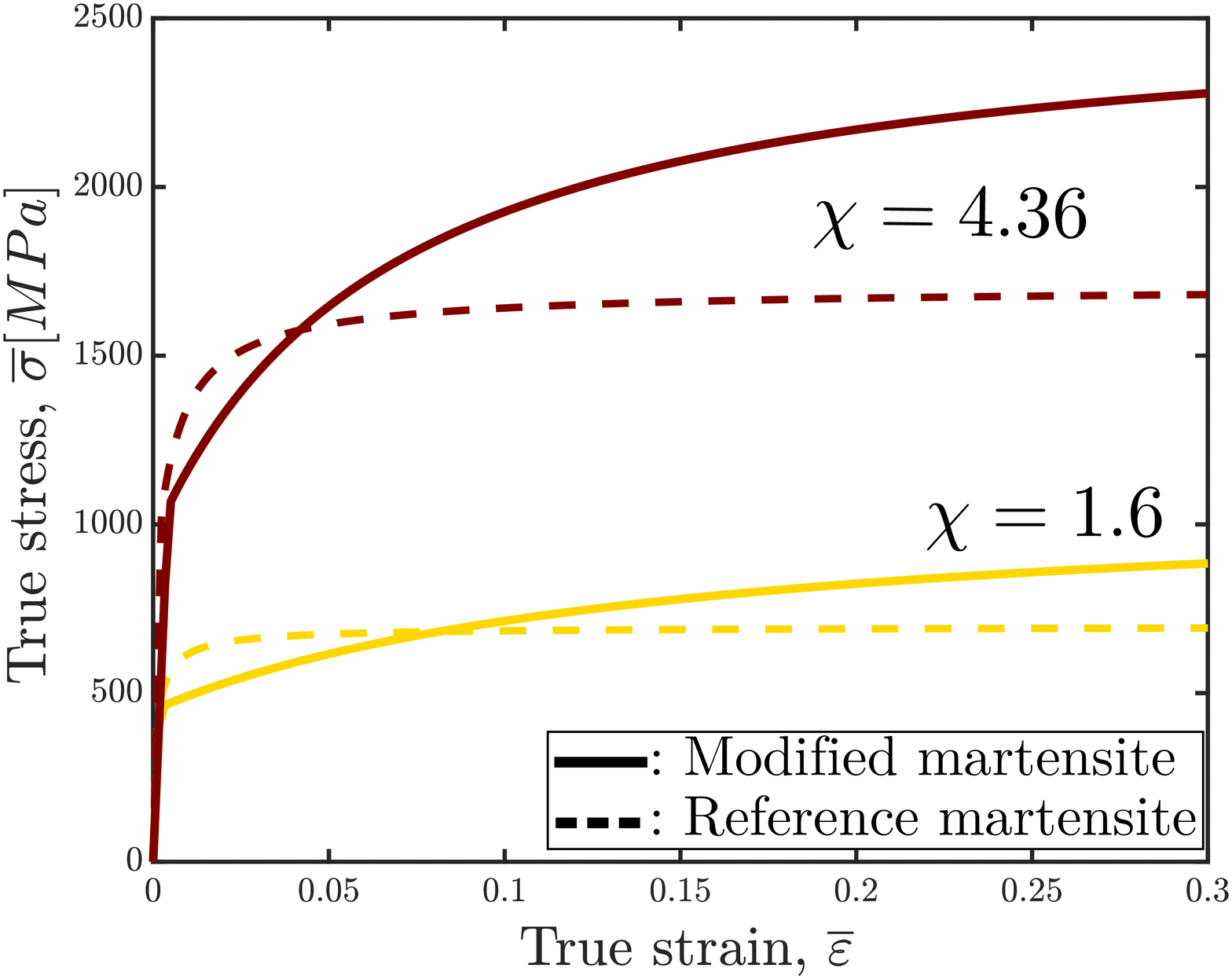}
\caption{Illustration of the modified martensite response, with more sustained hardening, as considered in Section \ref{SubSec:remedy}. The solid curves show the modified martensite phase response used in the simulations of the two new virtual steel grades. For these two grades, the lowest, $\chi=1.6$ (blue curves), and the highest, $\chi=4.36$ (red curves), mechanical phase contrast are considered. Other material parameters are fixed. The dashed curves are the reference martensite behavior considered so far. Note that the strain hardening capacity of the new martensite has increased by decreasing the initial hardening and increasing saturation stress value -- see Eqn. \ref{flowressistane}. }
\label{fig:NewMartensite}
\end{figure}

	New martensite material parameters, resulting in a higher hardening capacity, are adopted for the two virtual steel grades with the lowest and the highest mechanical phase contrast presented in Figure \ref{fig:PhaseBehaviors}. The hardening capacity of the martensite is increased by decreasing the initial hardening and increasing the saturation stress (to avoid very fast saturation). This allows the martensite grains to continue to harden during the whole deformation. With this approach, we keep the phase contrast (the ratio of the initial yield of two phases) and strength of the steels fixed, e.g. at $600$ MPa. Note that all the other parameters of the microstructure such as the ferrite behavior, martensite initial yield, and volume fraction remain unchanged. The comparison of the responses of the reference and modified martensite is shown in Figure \ref{fig:NewMartensite}. 

\begin{figure}[ht!]
\centering
\includegraphics[width=0.6\textwidth]{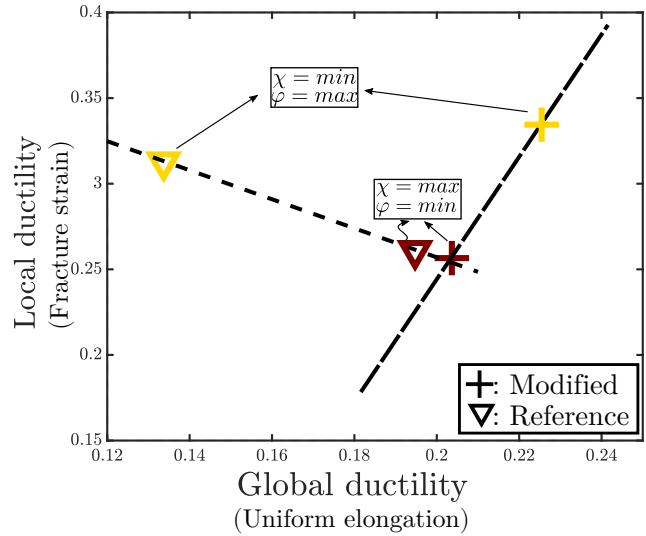}
\caption{Comparison of the global and local ductility of the two virtual grades with reference martensite behavior (see Figure \ref{fig:AFSteel_GL}) shown by ($\bigtriangledown$) markers, and modified martensite parameters, i.e. incorporating lower initial hardening and high strain hardening capacity at high strains, shown by ($+$) markers. The dashed curves indicates the trend of global and local ductility for both cases. For these two grades, the lowest, $\chi=1.6$, and the highest, $\chi=4.36$, mechanical phase contrast are considered.}
\label{fig:ReSteelAB}
\end{figure}

	Figure \ref{fig:ReSteelAB} compares the global and local ductility response of the two steels with the modified martensite behavior. Since the new parameter set for martensite allows it to harden more, the uniform elongation of both newly designed steel grades has increased -- cf. Figure \ref{fig:AFSteel_GL}. The increase in uniform elongation is higher in the steel with high martensite volume fraction $\varphi=0.55$. The local ductility of the steel with a low phase contrast, $\chi=1.6$, is improved only by $8\%$, whereas in the steel with a high phase contrast, $\chi=4.3$, no noticeable change in local ductility is observed. This is due to the fact that the high phase contrast in the microstructure results in damage mostly in ferrite and thus changing the martensite hardening does not affect the fracture strain in this case. It is shown that by considering the modified behavior of the martensite, which incorporates more hardening, the trade-off of local and global ductility has been removed: the microstructure with a low phase contrast, $\chi=1.6$, and a high martensite volume fraction, $\varphi=0.55$, performs better in terms of both the local and global ductility.

	The results reveal that to have a locally and globally ductile DP steel, increasing the strain hardening capacity of the martensite in higher strains can be a possible remedy. However, whether martensite behavior in DP steel can be tailored to have a favorable hardening behavior is indeed a challenge and the issue is not very straightforward to address. To our best knowledge, the origin of the very high initial hardening of martensite and subsequent early saturation has not been discussed in the literature.

		 To summarize, a DP steel which is to be used in a necking-controlled loading conditions, such as a tensile test, relies on a high mechanical phase contrast between its constituent phases that makes it globally necking-resistant, but locally (at the edges) damage-sensitive. Therefore, when a local ductility test such as a hole expansion test is conducted on this steel, the edges of the steel will fail earlier. Industry encounters this phenomenon as the 'cut-edge failure' or 'edge cracking' issue of DP steels \citep{Hoefnagels2020}. Yet, the problem is rooted in the fact that a DP steel designed for global ductility is not damage resistant. If cut-edge failure is to be improved, a DP steel with a higher martensite volume fraction and a lower mechanical phase contrast should be chosen instead, which in turn will reduce necking-controlled failure strain.

\section{Conclusion}
	Experimental findings in the literature reveal that DP steels with identical strength and composition, but different microstructures, exhibit inconsistent trends under either necking-controlled or damage-controlled ductility tests. A special case of this phenomenon is referred to in the forming community as the cut-edge failure or edge cracking issue of DP steels. It is observed that a globally ductile microstructures are prone to premature damage at cut edges while a comparatively less ductile DP steel performs better under the locally applied deformation at such edges. In this paper, a systematical statistical study is conducted to analyse this paradox from a micromechanical point of view. The obtained results qualitatively confirm the experimental observations and allowed to explain/rationalize them, as follows. To reach a higher necking-controlled ductility, i.e. global ductility, the mechanical contrast between ferrite and martensite must be increased. Since martensite reveals a very high initial hardening and fast saturation, activation of martensite plasticity will lead to early necking in the microstructure and should therefore be avoided; a high mechanical phase contrast has this effect. However, at the same time a higher mechanical phase contrast is detrimental for the local ductility. More heterogeneity in the microstructure causes the generation of higher local plastic strains and triaxial stresses which leads to early damages. Therefore, in DP steel, increasing the global ductility may come at the expense of decreasing local ductility, and vice versa. The martensite hardening behavior is the main factor governing this phenomenon, and therefore also constitutes the key in processing improved steels that combine the best of both.

\section*{Acknowledgements}
This research was carried out under project number T17019b in the framework of the Research Program of the Materials innovation institute (M2i) (\href{www.m2i.nl}{www.m2i.nl}) supported by the Dutch government.

\section*{Data availability statement}
The raw/processed data required to reproduce these findings cannot be shared at this time due to technical or time limitations.

\newpage
\bibliographystyle{unsrtnat}
\bibliography{/home/ador/Desktop/Papers/2ndPaper/paper/FinalVersion//ArXiv/main.bib}

\end{document}